\documentclass[12pt]{article}

\usepackage{esfconf}

\def\sla{\raise.15ex\hbox{$/$}\kern-.57em}

\def\Journal#1#2#3#4{{#1} {\bf #2}, #3 (#4)}

\def\NPB{{\em Nucl. Phys.} B}
\def\PLB{{\em Phys. Lett.}  B}
\def\PRL{\em Phys. Rev. Lett.}
\def\PRD{{\em Phys. Rev.} D}

\def\be{\begin{equation}}
\def\ee{\end{equation}}
\def\bea{\begin{eqnarray}}
\def\eea{\end{eqnarray}}

\begin{document}

\title{
Regulated chiral gauge theories.
\footnote{\em \tiny Invited talk at the international conference 
``Quantization, Gauge Theory, and Strings, dedicated to the memory
of Professor Efim Fradkin'',
June 5 - June 10, Moscow, Russia.}
}

\authors{H. Neuberger}

\addresses{Department of Physics and Astronomy, Rutgers University, 
Piscataway,\\ NJ 08855, USA\\E-mail: neuberg@physics.rutgers.edu}


\maketitle

\begin{abstract}
Noncompact chiral abelian gauge theories are defined on the lattice
using the overlap formalism. The main definitions are presented,
the role of anomaly cancelation is discussed, and the triviality
issue in four dimensions is explained. 
\end{abstract}

\vspace{-9.0cm}
\begin{flushright}{\normalsize RUHN-00-4}\\
\end{flushright}
\vspace{7.8cm}

\section{Introduction}

\vspace{-.3cm}

During the last several years we have learned how to
put chiral gauge theories on the lattice. For recent 
reviews and further reference see~\cite{tw}. The best
understood case is abelian and aims at defining the
theory on infinite Euclidean space $R^d$~\cite{nonc}.

We start by defining the Lagrangian. We work in even $d$-dimensional
Euclidean space,
on an infinite hypercubic lattice. The action has a pure gauge part
and a fermionic part. We shall start with the bosonic part, and next turn
to the fermionic component. This shall be followed by a display of the
lattice Wess Zumino action and of lattice anomalies. After that,
trivial cocycle components will be subtracted from the Wess Zumino
action leading to the non-perturbative 
restoration of gauge invariance when anomalies
cancel in perturbation theory. We then address the particular case of
$d=4$ where anomalous and non-anomalous theories are both trivial,
but in different ways. We conclude with a brief summary.

\vspace{-.2cm}
\section{Photons}
\vspace{-.3cm}

 The pure gauge term is given by:
\begin{equation}
S_{\rm pure ~gauge} =
{1\over {4e^2}} \sum_x F_{\mu\nu}^2  (x)
+{{m^2_\gamma}\over 2} \sum_x A_\mu^2 (x) +
{1\over {2\xi}} \sum_x [ \nabla_\mu^{x-} A_\mu (x) ]^2.
\end{equation}
The field strength is defined in terms of the noncompact
vector potential $A_\mu (x)$
by:
\begin{equation}
F_{\mu\nu} (x) = \nabla_\mu^{x+} A_\nu (x) - \nabla_\nu^{x+} A_\mu (x),
\end{equation}
where $\nabla_\mu^{x+}$ are forward finite difference operators
(the backward difference operators are $\nabla_\mu^{x-}=
(\nabla_\mu^{x+})^\dagger$). In
addition, gauge configurations have vanishing probability in the path
integral unless~\cite{bnd} for each site $x$ and each Lorentz index pair
$\mu\nu$, 
\begin{equation}
|F_{\mu\nu} (x) | \le \eta ,
\end{equation}
where,
\begin{equation}
0< \eta < {{1-(1+m)^2}\over {(1+{\sqrt{2}\over 2})d(d-1)}}.
\end{equation}
$m$ is a lattice parameter restricted by $|1+m| <1$ (for
our purposes here we can assume that $m$ has been fixed at $-1$).
Although we have added a photon mass term and a gauge
fixing term, gauge invariance plays a crucial role
in the quantization of the theory, the consistency
of its physical sector and the existence of a zero photon
mass limit. So long $m_\gamma^2 > 0$, $A_\mu (x)$ goes
to zero as $|x|\rightarrow \infty$ for any gauge field configuration 
of non-zero probability. The entire gauge invariant content of the
configuration is contained in the variables $F_{\mu\nu}(x)$.
On a finite lattice with translationally invariant boundary
conditions extra gauge invariant variables enter. These variables
cause complications that we avoid here. So long $0< \xi <\infty$
the gaussian path integral over $A_\mu (x)$ can be defined even
before the restriction of equation (3) is enforced. 

\section{Chiral Fermions}

The theory contains several Weyl fermions, either left or
right handed, with charges $q_1 , q_2 , ....q_n, 1$, $|q_i|\le 1$;
at least one of the fermions has $|q|=1$.  
For each fermion we define unitary matrices,
\begin{equation}
U_\mu (x;q) = e^{iqA_\mu (x)},
\end{equation}
and associated parallel transporters,
\begin{equation}
T_\mu (A) (\psi ) (x) = U_\mu (x) \psi (x+\hat\mu)
\end{equation}
acting on fermions $\psi$. The Wilson Dirac operator is given by:
\begin{equation}
D_W (A) = m+\sum_\mu(1- V_\mu );~~~V_\mu=
{{1-\gamma_\mu}\over 2} T_\mu +{{1+\gamma_\mu}\over 2} T_\mu^\dagger;
~~~V_\mu^{-1}=V_\mu^{\dagger}
\end{equation}
Associated to it, there is a hermitian Wilson Dirac operator
\begin{equation}
H_W(A)=\gamma_{d+1}D_W(A) .
\end{equation}
Here, $\gamma_{d+1}$ generalizes the Hermitian $\gamma_5$ matrix in four
dimensions. 

It is true then that the following unitary-hermitian operator~\cite{do} is
well defined for all gauge configurations
that can occur with non-zero probability:
\begin{equation}
\epsilon(A) ={\rm sign} (H_W (A)).
\end{equation}

To add the fermions we introduce an auxiliary Fock space \cite{ovlapop} 
generated by

\begin{equation}
\{ {\hat a}^\dagger_\alpha (x) , {\hat a}_\beta (y) \} = 
\delta_{\alpha\beta}
\delta_{xy}.
\end{equation}
In this Fock space we define a Hamiltonian:
\begin{equation}
{\hat H}(A)={\hat a}^\dagger \epsilon (A) {\hat a}+N_v, 
\end{equation}
where the constant $N_v$ assures that the ground state energy 
of ${\hat H}(A)$ be zero for all $A$.\footnote{Although the ground
state energy of ${\hat a}^\dagger \epsilon (A) {\hat a}$ is formally
infinite because of the infinite volume, it is a meaningful fact that
this energy is independent of $A$, for any $A$-background of non-zero 
probability. Moreover, all other eigenstates of ${\hat H}(A)$ have trivial,
integral, eigenvalues, which also do not depend on $A$.} 
There are local charge operators given by:
\begin{equation}
{\hat n}(x) = {1\over 2} 
[{\hat a}^\dagger (x) {\hat a}(x) - 
{\hat a}(x){\hat a}^\dagger (x)].
\end{equation}
The total charge operator 
${\hat N} =\sum_x {\hat n}(x)$ commutes with the Hamiltonian.
The ground state of ${\hat H}(A)$ is unique and of zero
charge for all $A$:
\begin{equation}
{\hat H}(A)| v(A) \rangle =0,~~~~{\hat N} |v(A) \rangle =0 .
\end{equation}
Gauge transformations are generated by the exponentiation
of the local charge operators. The resulting unitary
operators are given by:
\begin{equation}
{\hat G} (\alpha ) = e^{i\sum_x \alpha (x) {\hat n}(x)}.
\end{equation}
The Hamiltonian transforms covariantly under a
gauge transformation:
\begin{equation}
{\hat G} (\alpha ) {\hat H}(A) 
{\hat G}^\dagger (\alpha ) = {\hat H} 
(A^{(\alpha )} ).
\end{equation}
Here, $A^{(\alpha)}_\mu (x) \equiv A_\mu (x) +\nabla_\mu^{x+} \alpha(x) 
\equiv A_\mu (x) + \alpha (x+\mu ) - \alpha (x)$,
is a gauge transform of $A_\mu (x)$.  

We also define a reference system, with $\epsilon(A)\rightarrow \gamma_{d+1}$, 
${\hat H} (A)\rightarrow {\hat H}^\prime$ and $|v (A)\rangle\rightarrow
|v^\prime \rangle$. Trivially, ${\hat G} (\alpha )|v^\prime \rangle =
|v^\prime \rangle$. The phase of $|v^\prime \rangle$ is fixed in an arbitrary
way. The phase of $|v(A)\rangle$ is fixed by an adiabatic phase choice:
\begin{equation}
\langle  v(tA )|{ {dv (tA )} \over{dt}}  \rangle =0;~~~
\langle v^\prime | v(0) \rangle > 0. 
\end{equation}

The overlap provides a finite formula
for the fermionic determinant: 
\begin{equation}
\langle v^\prime | v(A)\rangle .
\end{equation}
This formula can be motivated by considering a system containing an
infinite number of Dirac fermions with a mass matrix of the order of
the ultraviolet cutoff, but with unit index. This interpretation connects
the formula to the basic papers of Kaplan~\cite{kap} 
and of Frolov and Slavnov~\cite{fro}. 

\section{Wess-Zumino action}

If we made no phase choices, the overlap could be viewed as a line bundle
over the space of gauge configurations. 
Because of the gauge transformation
properties of the two states, this bundle descends into a line bundle
over the space of gauge orbits. The phase choice can be viewed as a choice
of section in the former, but, generically, 
not in the latter because the section
may vary along gauge orbits~\cite{geom}. 
Lack of ordinary gauge invariance is measured
by the Wess-Zumino action $\Phi (\alpha , A)$ 
which is given by
\begin{equation}
\langle v^\prime | v(A^{(\alpha)})\rangle = e^{i\Phi(\alpha, A)}
\langle v^\prime | v(A)\rangle .
\end{equation}
Starting from the adiabatic 
phase choice, the Wess Zumino term can be explicitly calculated:
\begin{equation}
\Phi (\alpha , A) = -\sum_x \alpha (x) \int_0^1 dt
\langle v (tA)|{\hat n}(x) | v(tA )\rangle .
\end{equation}
$\Phi(\alpha, A)$ is linear in $\alpha$ and vanishes for all $\alpha$
if $A=0$. 
The factor multiplying $\alpha(x)$ is the local charge
density and sums over the lattice to zero, 
for any $A$-configuration that occurs with non-zero
probability.\footnote{It is amusing to note here that in earlier
attempts to put chiral fermions on the lattice it was assumed
that the decoupling of $\alpha(x)$ on the trivial orbit was the
central problem one was facing on the lattice. We now see that the
lattice is not different from the continuum in this respect and
whether the theory is anomalous or not, the gauge degrees of freedom
can be made to decouple on the trivial orbit. This has no bearing
on the existence of a continuum limit, which does depend on anomaly
cancelation. Thus, as one should have known from the continuum, the
main obstacles to the quantization of a chiral gauge theory on the 
lattice cannot be even identified before fermion loops are taken into
account. However, the majority of previous attempts to put chiral fermions
on the lattice used various forms of Yukawa type interactions
and were deemed unsuccessful even before any fermion loop was taken
into account.}  

\section{Anomalies}

The Bose symmetrized (``consistent'') anomaly comes directly from $\Phi$:
\begin{equation}
\triangle_{\rm cons} (x;A) 
= -\int_0^1 dt \langle v(tA)|{\hat n}(x)
|v(tA)\rangle .
\end{equation}
One easily establishes that
\begin{equation}
\triangle_{\rm cons} (x;A)=-\nabla_\mu^{x-} {\cal J}_{\mu \ {\rm
cons}}(x;A);~~~
{\cal J}_{\mu \ {\rm cons}}
(x;A) = {{\partial \log \langle v^\prime | v (A) \rangle}\over
{\partial A_\mu (x)}} .
\end{equation}
In addition to the consistent current we have the covariant 
current~\cite{geom}
\begin{equation}
{\cal J}_{\mu \ {\rm cov}} (x;A) = {1\over
{ \langle v^\prime | v(A) \rangle}}
\langle v^\prime |{{ \partial v (A) }\over
{\partial A_\mu (x)}}\rangle_\perp ,
\end{equation}
with a covariant anomaly:
\begin{equation}
\triangle_{\rm cov} (x;A)
=-\nabla_\mu^{x-} {\cal J}_{\mu \ {\rm
cov}}(x;A) =
\langle v(A)|{\hat n}(x)|v(A)\rangle .
\end{equation}
These currents are related to a current operator in Fock space,
\begin{equation}
[{\hat n}(x), {\hat H}(A)]= i \nabla_\mu^{x-} {\hat J}_\mu (x;A),
\end{equation}
which admits the form:
\begin{equation}
{\hat J}_\mu (x;A) = -{{\partial {\hat H} (A)}
\over {\partial A_\mu (x) }} .
\end{equation}
The relation to the c-number currents above is:
\begin{equation}
\langle v(A) | {\hat n }(x) | v(A) \rangle = 
\nabla_\mu^{x-} j_\mu (x; A),
\end{equation}
where
\begin{equation}
j_\mu (x; A)= \int_0^1 dt
{i\over 2}
\left [ \langle v(tA) | {\hat J}_\mu (x;tA) 
| {{dv(tA)}\over {dt}}\rangle -c.c.\right ].
\end{equation}
$j_\mu (x; A)$ is not gauge invariant:
\begin{equation}
j_\mu (x;A^{(\alpha )}) - j_\mu (x; A)= 
\end{equation}
\begin{equation}
{i\over 2} \sum_y  \left [
\int_0^1 dt \langle v(tA)| [ {\hat J}_\mu (x;tA), {\hat J}_\nu (y;tA)]|v(tA)
\rangle \right ] \nabla_\nu^{y+} \alpha (y) 
\end{equation}
This introduces the Schwinger term~\cite{nonc}:
\begin{equation}
S_{\mu\nu} (x,y;A)= \langle v(A)| [ {\hat J}_\mu (x;A), {\hat
J}_\nu
(y;A)]|v(A)\rangle .
\end{equation}
Only the divergence of the Schwinger term enters in equation (29). 
The matrix element of the current commutator turns out to be  
Berry's curvature associated with the line
bundle $|v (A)\rangle$, where $A$ is viewed
as ``parameter space''. (Note that this
line bundle does not descend to a bundle over
gauge orbits, but, its Berry curvature can
be viewed as a two form over the space of
gauge orbits.) Let us present the explicit
expressions. The Berry connection is defined by
\begin{equation}
{\cal A}_{\mu x} (A) = \langle v(A) | {{\partial v(A)}\over
{\partial A_\mu (x) }} \rangle,
\end{equation}
from which we get Berry's curvature~\cite{geom}:
\begin{equation}
{\cal F}_{\mu x, \nu y} (A) = {{\partial {\cal A}_{\mu x} (A)}
\over {\partial A_\nu (y) }} - {{\partial {\cal A}_{\nu y} (A)}
\over {\partial A_\mu (x) }}=
\end{equation}
\begin{equation}
\langle {{\partial v(A)}\over {\partial A_\mu (x)}} | 
{{\partial v(A)}\over {\partial A_\nu (y) }} \rangle -
\langle {{\partial v(A)}\over {\partial A_\nu (y)}} | 
{{\partial v(A)}\over {\partial A_\mu (x) }} \rangle .
\end{equation}
One can easily show that~\cite{nonc}:
\begin{equation}
S_{\mu\nu} (x,y;A)=4{\cal F}_{\mu x, \nu y} (A).
\end{equation}
The c-number 
covariant current can also be related to a matrix 
element of the current operator:
\begin{equation}
{\cal J}_{\mu \ {\rm cov}}(x;A)=-{1\over 2}
{{\langle v^\prime | {\hat J}_\mu (x; A) |v(A)\rangle}\over
{ \langle v^\prime | v(A) \rangle}} .
\end{equation}
Both the covariant and consistent currents involve, in addition
to $|v(A)\rangle$, also the reference state and are non-local
functionals of $A$. The difference between the currents, given by Berry's
curvature, does not involve $|v^\prime\rangle$, and is a local
funtional of $A$. 
\section{Trivial cocycles}

One can establish, by explicit construction or by more indirect
means, the following fact: 
If the continuum anomaly cancelation
condition is met, the lattice Wess-Zumino functional is a
locally trivial cocycle. To see what this means let us focus on two
dimensions as an example~\cite{nonc}. Define a constant $b$:
\begin{equation}
b= {1\over 2} \epsilon_{\mu\nu} \sum_x S_{\mu\nu} (x,y;0) .
\end{equation}
One can prove that there exist local gauge field functionals $\chi_\rho$, 
such that the divergence of the Schwinger
term can be written as:
\begin{equation}
\nabla_\nu^{y-} S_{\mu\nu} (x,y;A) = 
\epsilon_{\mu\sigma}\nabla_\sigma^{x-}[\nabla_\rho^{y-}\chi_\rho (x,y;A)
-b\delta_{x,y}].
\end{equation}
This reflects the 
identity $\nabla_\mu^{x-}\nabla_\nu^{y-} S_{\mu\nu} (x,y;A)=0$ which
holds as a result of equations (24) and (30). 
Using $\chi_\rho$, the adiabatic phase choice of $|v(A)\rangle$ can
be augmented by adding to the adiabaticity requirement an 
explicit phase factor given by the exponent of a local, non-gauge
invariant, functional of $A_\mu (x)$
\begin{equation}
|v(A)\rangle\rightarrow e^{i\phi(A)} |v(A)\rangle ,
\end{equation}
which changes the Wess Zumino functional by a trivial cocycle,
$\Phi (\alpha ,A)\rightarrow \Phi (\alpha ,A) +
\phi(A^{(\alpha )}) - \phi(A)$, so that the new Wess Zumino functional
becomes:
\begin{equation}
\Phi^{\rm new} (\alpha, A) = -{b\over 2} \sum_x F_{12} (x) \alpha(x) .
\end{equation}
For the opposite handedness the overall sign gets reversed. It is
now obvious that when the continuum anomaly cancelation condition
$\sum q_R^2 =\sum q_L^2$ is met the new Wess Zumino functional
vanishes, showing that the previous Wess
Zumino functional was a trivial cocycle. 
So, when anomalies cancel in the perturbative sense, gauge invariance
gets restored also outside perturbation theory. 

\section{Triviality}

While two dimensional chiral abelian gauge theories 
have a coupling dependent continuum
limit\footnote{
Chiral two dimensional models, for compact $U(1)$, on a finite torus,
have been studied in the overlap formalism in~\cite{twod}. The
formulation there 
is somehwat different from the one presented here, so that it be well
adapted to numerical simulation. Using it, it was shown that such subtle
phenomena as the appearance of massless fermionic composite particles,
just so that 't Hooft's consistency relations be satisfied, indeed
takes place for properly defined lattice models.}
, four dimensional chiral abelian gauge
theories are ``trivial'' because 
the physical coupling is forced
to vanish if one takes the ultraviolet cutoff to infinity. An anomaly
free chiral $U(1)$ theory, just like QED or the Higgs sector~\cite{higgs}
of the minimal standard model, does not, strictly
speaking, have an interacting continuum limit. Since the continuum limit
is also jeopardized by lack of anomaly cancelation it is necessary to
understand the difference between the two kinds of 
``trivialities'' in four dimensions~\cite{nonc}.

In four dimensions it is not necessary
to deal with both left and right Weyl fermions,
because a left handed fermion can
be viewed as a right handed one in the conjugate
representation. The anomaly cancelation condition can be written
then as $\sum_{\rm left~ handed~ fermions} q^3 =0$; 
if it is not satisfied, there is an obstruction
to finding a section in the compounded bundle over $A$ that descends
into a section over gauge orbits~\cite{geom}. This generalizes
to the non-abelian case~\cite{adams}. 

The difference between the two cases of
triviality can be expressed by employing the effective field theory
framework. Whether anomalies cancel or not, we can write down a theory
which interacts but also has an ultraviolet cutoff $\Lambda$. This theory
is adequate to describe to a given finite accuracy processes of
energy up to some physical scale $E_{\rm ph}$. Clearly, 
$E_{\rm ph}<\Lambda$. For the accuracy of this description to be sensible
the physical coupling $e^{\rm ph}$ cannot be too large.
If anomalies do not cancel the restriction has the structure:
\begin{equation}
e^{\rm ph} \le c_1 \left ( {{E^{\rm ph}}\over \Lambda} \right )^{1\over 3},
\end{equation}
where $c_1$ is some constant.
If anomalies do cancel the restriction is
\begin{equation}
e^{\rm ph} \le {c_2\over{\log \left (  {\Lambda\over{E^{\rm ph}}} \right )}} .
\end{equation}
So, in both cases one cannot take $\Lambda$ to infinity while
keeping $e^{\rm ph}$ non-zero, but the restriction is much weaker
if anomalies cancel. 

\section{Summary}

There is
little doubt left that indeed, as anticipated by most particle
physics theorists, anomaly free, asymptotically free four dimensional
chiral gauge theories have consistent continuum limits and exist
outside perturbation theory. These are non-trivial theories, with no
free parameters. These theories
play an important role in our understanding of Nature, and what
kind of mechanisms make them appear naturally at  
energies presently accessible 
is an important issue which may hold
some hints about how things work in the real world.

\section*{Acknowledgments}
This research was 
supported in part by the DOE under grant \#
DE-FG05-96ER40559. I wish to thank the organizers of the ``Quantization,
Gauge Theory and Strings'' dedicated to the memory of Professor Efim Fradkin
for the invitation to participate, and for the hospitality
extended.

\end{document}